\documentclass[aps,prb,reprint]{revtex4-2}
\usepackage{hyperref}
\hypersetup{
    colorlinks = true,
    linkcolor = [rgb]{0.70, 0.13, 0.13},
    citecolor = [rgb]{0.13, 0.55, 0.13},
    urlcolor = [rgb]{0.25, 0.41, 0.88}
}
\usepackage{amsmath}
\usepackage{amsfonts}
\usepackage{amssymb}
\usepackage{pifont}
\usepackage{mathtools}
\usepackage{bm}
\usepackage{cases}
\usepackage{braket}
\usepackage[version=4]{mhchem}
\usepackage{graphicx}
\allowdisplaybreaks%

\DeclareMathOperator{\Tr}{Tr}

\begin{document}
\title{Intertwining SU($N$) symmetry and frustration on a honeycomb lattice}

\author{Xu-Ping Yao}
\thanks{These authors contributed equally.}
\author{Rui Leonard Luo}
\thanks{These authors contributed equally.}
\author{Gang Chen}
\email{gangchen@hku.hk}
\affiliation{Department of Physics and HKU-UCAS Joint Institute 
for Theoretical and Computational Physics at Hong Kong,
The University of Hong Kong, Hong Kong, China}
 
\date{\today}

\begin{abstract}
Large symmetry groups in quantum many-body systems 
could strongly enhance quantum fluctuations and thereby
stabilize exotic quantum phases. Frustrated interactions
were long known to have similar effects. Here we intertwine 
the large SU($N$) symmetry and the frustration in a $J_1$-$J_2$ 
SU($N$) Heisenberg model on a honeycomb lattice, where $J_1$ 
is the nearest-neighbor coupling and $J_2$ is the next-nearest-neighbor 
coupling. With a large-$N$ analysis, we obtain a rich phase diagram 
by varying both $N$ and the ratio $J_2/J_1$. 
The ground states include Dirac spin liquid, chiral spin liquid, 
valence cluster solids, flux ordered state, and stripe states. 
The physical properties of each phase are discussed.
\end{abstract}


\maketitle

\section{Introduction}
\label{Sec:Introduction}

In quantum many-body systems with a large symmetry group, 
the quantum fluctuations can be intensively enhanced and thus 
prevent the formation of the conventional orders. 
Therefore, the large symmetry group provides an interesting 
direction to stabilize novel and exotic quantum states. 
This scenario is fundamentally different from the common classical limit 
with a large spin moment in most solid-state magnets where a large local 
Hilbert space is also encountered. Over there, the model Hamiltonian 
does not have nor is proximate to a large symmetry group to access the   
large local Hilbert space effectively to enhance the quantum fluctuations. 
Therefore, having or being proximate to a large symmetry group can be one 
important ingredient to realize exotic quantum phases. 
In condensed matter physics, the large symmetry group is often 
considered as a theoretical fantasy to access exotic quantum phases
and limits~\cite{Sachdev_1991,PhysRev.185.847,Auerbach1994}.

Now several realistic quantum many-body systems may turn
this theoretical fantasy into reality. As a representative, 
the ultracold-atom system has been substantially developed 
to achieve the large symmetry like the SU($N$) symmetries in 
the fermionic cold gases, especially 
alkaline-earth atoms (AEAs)~\cite{10.1038/nphys1535, 10.1038/nphys2430,
PhysRevLett.91.186402,PhysRevLett.95.266404,PhysRevLett.112.156403}. 
The past few years have witnessed the reports of many nontrivial 
phenomena in this platform including Mott crossover, antiferromagnetic spin correlation, 
bosonization of the SU($N$) fermions, Pomeranchuk effects, and pronounced interaction 
effects~\cite{10.1038/nphys3061, PhysRevX.6.021030, PhysRevX.10.041053, PhysRevLett.121.225303, taie2020observation, 10.1038/s41586-018-0661-6,10.1038/nphys2430}. 
Quite recently, the highly tunable two-dimensional (2D) moir\'{e} materials have been proposed 
to be a new candidate where the Hubbard models with SU(4) and SU(8) 
symmetries can be realized through meticulously designed stacking 
and twisting~\cite{PhysRevLett.121.087001, zhang2020electrical, PhysRevLett.127.247701}.

In this rapidly evolving field, the SU($N$) Mott insulators have attracted significant 
attention because they are a straightforward generalization of the conventional 
SU(2) one. Tremendous efforts have been made on the theoretical side to reveal 
their nature and it turns out that various interesting ground states could emerge 
depending not only on the lattice and the number $N$ but also on the filling per site~\cite{PhysRevLett.103.135301, PhysRevB.84.174441, PhysRevResearch.3.023138, 
chen2021abelian, PhysRevA.84.043601, PhysRevA.93.061601, PhysRevB.71.075103, 
PhysRevB.97.064415, PhysRevB.97.064416, PhysRevB.96.205142,PhysRevLett.121.225303}. 
Although the SU($N$) symmetry seems to be more like an idealization 
in the realistic solid-state materials than the ultracold atom systems, 
the Hubbard model with emergent SU(4) symmetry has been 
proposed to capture the effective physics of the spin-orbital 
compounds such as the Kugel-Khomskii spin-orbital 
system \ce{Ba3CuSb2O9}~\cite{PhysRevB.100.205131} 
and the spin-orbit-entangled system 
$\alpha$-\ce{ZrCl3}~\cite{PhysRevLett.121.097201, PhysRevB.104.224436}, 
and even on the moir\'{e} superlattice 
such as twisted bilayer graphene and transition metal dichalcogenides~\cite{PhysRevLett.121.087001, zhang2020electrical, PhysRevLett.127.247701}.

\begin{figure}[b]
    \centering
    \includegraphics[width=8.6cm]{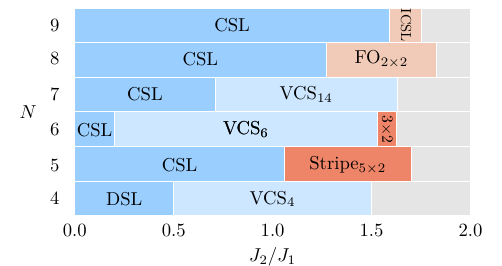}
    \caption{(Color online.) Phase diagram for the SU($N$) spins (${4 \le N \le 9}$) 
    with $J_1$-$J_2$ Heisenberg interactions on the honeycomb lattice. 
    The CSL and inhomogeneous chiral spin liquid (ICSL) states have total flux $4\pi/N$ and $2\pi/N$ 
    through the unit cell, respectively. The subscripts for valence cluster solids (VCSs) represent 
    the numbers of sites in each cluster. The subscripts for stripe and flux ordered (FO) 
    states indicate the periods of the enlarged unit cells. In the gray regions, 
    the system is decoupled into two equivalent triangular subsystems.}
    \label{Fig:PhaseDiagram}
\end{figure}


In particular, the SU($N$) spin physics on the honeycomb lattice 
with only the nearest-neighbor antiferromagnetic Heisenberg interactions 
has been extensively studied from both theoretical and numerical perspectives. 
For the SU(2) spins, the system is unfrustrated 
due to the bipartiteness of the lattice and the ground state 
has an antiferromagnetic N\'{e}el order. 
A valence bond state with the hexagonal plaquettes is found 
 for the SU(3) spins by tensor network simulations 
and further confirmed by exact diagonalizations 
and variational Monte Carlo studies~\cite{PhysRevB.85.134416,PhysRevB.87.195113,PhysRevB.100.035134}. 
The SU(4) case is the most attractive because it is believed to support 
a peculiar spin liquid state known as the U(1) Dirac spin liquid 
(DSL)~\cite{PhysRevX.2.041013}.  In contrast, the ${N = 5}$ case is less studied.
The ground states of SU(${ N \ge 6}$) spins are inferred to be chiral spin liquid (CSL) 
states~\cite{PhysRevA.84.011611,PhysRevA.88.043619,PhysRevLett.103.135301} 
albeit the SU(6) case is still under debate~\cite{PhysRevB.93.201113}.

Another important but more well-known ingredient to stabilize 
exotic quantum phases is the magnetic frustration~\cite{BalentsNature}. 
For the honeycomb lattice that is discussed here, the introduction
of the next-nearest-neighbor Heisenberg interactions
would frustrate the antiferromagnetic N\'{e}el ground state
of the nearest-neighbor SU(2) Heisenberg interaction. 
Even at the classical level, it has been shown that a spiral spin liquid 
regime could emerge from the N\'{e}el-ordered state in the large-$S$ 
(classical) limit where quantum fluctuations are suppressed~\cite{PhysRevB.81.214419,10.1007/s11467-021-1074-9dzm}.
 For the spin-1/2 quantum case,
 the density matrix renormalization group calculation 
suggests a deconfined quantum phase transition from the N\'{e}el 
order to the plaquette order~\cite{PhysRevLett.110.127203}.


In this work, we plan to explore the interesting situation  
where both the large symmetry group and frustration are 
present in one system, and examine the consequences by 
intertwining these two ingredients. 
The SU($N$) honeycomb lattice $J_1$-$J_2$ spin model is 
a manifestation of this intertwining. Despite the existing results on
the nearest-neighbor SU($N$) honeycomb lattice spin model, 
the role of the next-nearest-neighbor 
interactions for general SU($N$) spins with enhanced quantum fluctuations 
is not yet clear except for ${N \le 3}$~\cite{PhysRevLett.107.087204, PhysRevB.88.165138, 
PhysRevB.92.195110, PhysRevB.93.214438}. 
Moreover, the low-energy physics of the U(1) DSL for the SU(4) spins 
is effectively captured by a compact QED$_3$ theory with ${N_f = 8}$ 
Dirac fermions coupled to a dynamic U(1) gauge field~\cite{PhysRevB.70.214437}.
It is generally believed that the large number of gapless matter could
stabilize the U(1) DSL by suppressing the space-time monopole events.
It is, however, still unclear about the lower bound for the critical number
of the gapless matter modes for this stabilization~\cite{PhysRevLett.112.151601,calvera2021theory}.
Since the U(1) DSL is expected to be the parent state of many competing states 
in two dimensions by the spontaneous generation of the Dirac masses~\cite{Song_2019}, 
it would be interesting to explore the nearby (descending) phases 
with the SU($N$) spin systems. Therefore, the fate of the U(1) DSL 
with further interactions has triggered growing research 
interests~\cite{PhysRevB.100.241111, dupuis2021anomalous, 
PhysRevB.104.075103, iqbal2021gutzwillerprojected}.
We attempt to fill the gap at the mean-field level 
by investigating the SU(${4 \le N \le 9}$) Heisenberg model on the 
honeycomb lattice with both nearest-neighbor and 
next-nearest-neighbor antiferromagnetic interactions. 
Given that the decoupled limit where the system reduces to the triangular 
antiferromagnetic model has been explored in our previous 
work~\cite{PhysRevResearch.3.023138}, we focus on the finite next-nearest-neighbor 
interactions 
here and construct the phase diagram in the large-$N$ approximation. 
The mean-field results are summarized in Fig.~\ref{Fig:PhaseDiagram}. 
Our results confirm the existence of the putative DSL state for SU(4) spins 
and the CSL states for larger parameter $N$. 
It is found that both of them have a uniform background U(1) gauge flux 
$4\pi/N$ piercing the hexagonal plaquettes and remain stable against the
weak next-nearest-neighbor interactions. When the next-nearest-neighbor interactions become stronger 
but have not driven the system into the decoupled regime, 
a plethora of intermediate 
quantum states emerge for different parameter $N$,
 including the inhomogeneous chiral spin liquids (ICSLs), valence cluster solids 
(VCSs), and stripe and flux ordered (FO) states. 
All these intermediate quantum states break various lattice translational symmetries. 
The richness of the phase diagram reveals the intense competition of
the low-energy states due to the quantum fluctuation and frustration.

The rest of the paper is organized as follows. 
The SU($N$) Hubbard model and the derivative SU($N$) 
Heisenberg Hamiltonian in the strong-coupling limit are introduced 
in Sec.~\ref{Sec:Model}. With the representation of the constrained fermions, 
a mean-field Hamiltonian is obtained in the large-$N$ limit,   
whose parameters are defined by the saddle-point equations.    
Then a self-consistent minimization algorithm
is employed to solve the saddle-point equations and find the ground  
states of the mean-field Hamiltonian strictly satisfying the local constraints. 
Specifically, the DSL and the descending tetramer states for the SU(4) spins 
are discussed in Sec.~\ref{Sec:Results2-4}. 
The CSL and other intermediate quantum phases 
for higher SU($N$) spins are described in Sec.~\ref{Sec:Results5-9}. 
The paper is concluded in Sec.~\ref{Sec:Conclusion}.

\section{Large-$N$ approximation of SU($N$) Heisenberg model}
\label{Sec:Model}

The SU($N$) Hubbard model at $1/N$ filling 
(or equivalently with one particle per site) can be reduced 
to the Heisenberg model of SU($N$) spins in the strong-coupling limit 
up to second order. 
The effective spin with an internal SU($N$) symmetry 
is naturally introduced at each lattice site and can be 
expressed with the $N$-flavor Abrikosov fermions as 
${S_{\alpha\beta}(\bm{r})=f_{\bm{r}\alpha}^{\dagger}f_{\bm{r}\beta}}$ 
where ${\alpha, \beta = 1, \ldots, N}$.
This fundamental representation is accompanied 
by a local constraint on the fermions 
${f_{\bm{r}\alpha}^{\dagger}f_{\bm{r}\alpha} = 1}$ 
to reduce the enlarged Hilbert space. 
Note that a summation over repeated flavor indices 
is supposed hereafter unless otherwise specified.
We consider such an SU($N$) Heisenberg model 
on the honeycomb lattice,
\begin{equation}
\label{Eq:SpinH}
    \mathcal{H} = 
    J_1 \sum_{\braket{\bm{r}\bm{r}'}}S_{\alpha\beta}(\bm{r})S_{\beta\alpha}(\bm{r}') 
    + J_2 \sum_{\braket{\braket{\bm{r}\bm{r}'}}}S_{\alpha\beta}(\bm{r})S_{\beta\alpha}(\bm{r}'),
\end{equation}
where both the nearest-neighbor and next-nearest-neighbor exchange interactions $J_{1,2}$ are antiferromagnetic. 
It is apparent that in the ${J_2/J_1 \rightarrow \infty}$ limit, this model is
equivalent to two decoupled SU($N$) Heisenberg models on triangular sublattices 
that have been studied in our previous work and others~\cite{PhysRevResearch.3.023138,Keselman_2020,Keselman_2020b,jin2021unified}.
Here we still employ the large-$N$ saddle-point approximation to explore the nature in the moderate $J_2/J_1$ regime for different SU($N$) spins. 
Distinct from a perturbative expansion in the size of the interactions, this method has the advantages of preserving the spin symmetry and controlling systematic error by the higher-order correction in $1/N$~\cite{Auerbach1994,PhysRevB.84.174441}.
The partition function of the spin Hamiltonian in Eq.~\eqref{Eq:SpinH} can be expressed in the form of an imaginary-time functional integral:
\begin{equation}
    \mathcal{Z} = \int \mathcal{D}\chi^{\dagger} \mathcal{D}\chi \mathcal{D}\mu \mathcal{D}f^{\dagger} \mathcal{D}f e^{- \mathcal{S}}.
\end{equation}
The action is given as
\begin{multline}
    \mathcal{S} = \int_{0}^{\beta}d\tau \bigg\{ \sum_{\bm{r}}  f_{\bm{r}\alpha}^{\dagger} \partial_{\tau} f_{\bm{r}\alpha}^{} + \mu_{\bm{r}}(f_{\bm{r}\alpha}^{\dagger}f_{\bm{r}\alpha}^{} - 1)  \\
 \,\,  + \sum_{\braket{\bm{r}\bm{r}'},\braket{\braket{\bm{r}\bm{r}'}}} (\chi_{\bm{r}\bm{r}'}^{} f_{\bm{r}\alpha}^{\dagger}f_{\bm{r}'\alpha}^{} + \text{H.c.}) + \frac{N}{\mathcal{J}_{\bm{r}\bm{r}'}} |\chi_{\bm{r}\bm{r}'}^{}|^2 \bigg\}.
\end{multline}
A set of Lagrangian multipliers $\mu_{\bm{r}}$ has been introduced 
to enforce the single occupation constraint on each lattice site. 
There are also two types of auxiliary fields $\chi_{\bm{r}\bm{r}'}$ for
the nearest-neighbor and next-nearest-neighbor bonds, to decouple the fermion operators.
For the sake of simplicity, we have redefined the exchange couplings 
as ${\mathcal{J}_{\braket{\bm{r}\bm{r}'}} = N J_1 = 1}$ and 
${\mathcal{J}_{\braket{\braket{\bm{r}\bm{r}'}}} = N J_2 = J_2/J_1}$.
Taking the large-$N$ limit on the action $\mathcal{S}$ 
leads to the mean-field Hamiltonian for the noninteracting fermionic spinons,
\begin{eqnarray}
\label{Eq:HMF}
    \mathcal{H}_{\text{MF}} & = & \sum_{\braket{\bm{r}\bm{r}'},\braket{\braket{\bm{r}\bm{r}'}}} \frac{N}{\mathcal{J}_{\bm{r}\bm{r}'}} |\chi_{\bm{r}\bm{r}'}^{} |^2 
    + (\chi_{\bm{r}\bm{r}'}^{} f_{\bm{r}\alpha}^{\dagger}f_{\bm{r}'\alpha}^{} +\text{H.c.}) 
 \nonumber    
    \\
 && \quad\quad    + \sum_{\bm{r}}\mu_{\bm{r}}(1 - f_{\bm{r}\alpha}^{\dagger} f^{}_{\bm{r}\alpha}),
\end{eqnarray}
and the saddle-point equations
\begin{eqnarray}
  &&  \braket{f_{\bm{r}\alpha}^{\dagger} f_{\bm{r}\alpha}^{}} = 1, \label{eq:local_constraint} \\
    &&\chi_{\braket{ \bm{r}\bm{r}' }} = 
    - \braket{ f_{\bm{r}'\alpha}^{\dagger} f_{\bm{r}\alpha}^{} }/N, \\ 
&&    \chi_{\braket{\braket{ \bm{r},\bm{r}' }}} 
=  - J_2/J_1 \braket{ f_{\bm{r}'\alpha}^{\dagger} f_{\bm{r}\alpha}^{} }/N. 
\end{eqnarray}

In the following sections, we determine the ground-state phase diagram of 
the spinon mean-field Hamiltonian in Eq.~\eqref{Eq:HMF} with ${2\le N \le 9}$ 
numerically by utilizing the self-consistent minimization (SCM) algorithm 
developed in Refs.~\cite{PhysRevLett.103.135301,PhysRevB.84.174441}.
We extend this algorithm in order to involve the antiferromagnetic next-nearest-neighbor 
interactions by treating the two types of auxiliary fields $\chi_{\bm{r}\bm{r}'}$ 
on the honeycomb lattice and triangular sublattices synchronously 
and updating the chemical potentials $\mu_{\bm{r}}$ unitedly.
The rest of the technical details are briefly described in the Appendix. 
It should be emphasized that, differing from the usual analytical method 
where the local constraints are enforced only on average, 
the numerical SCM algorithm faithfully respects the single 
occupation constraints on each lattice site~\cite{PhysRevB.84.174441}.
Therefore, the obtained results are very reliable at the mean-field level, 
especially when the systematic correction beyond
the mean-field results becomes negligible with 
increasing the flavor number $N$.

\begin{figure}[t]
    \centering
    \includegraphics[width=8.6cm]{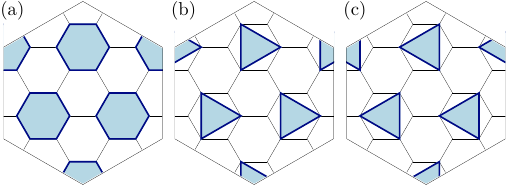}
    \caption{(Color online.) The VCS states for the SU(3) spins. (a) An ordered hexagonal pattern form on the nearest-neighbor bonds coexisting with (b, c) the three-site simplex VCS patterns on the next-nearest-neighbor bonds. The expectation values $\chi_{\braket{\bm{r}\bm{r}'}}$ decrease with the increasing nearest-neighbor exchange interaction and eventually vanish at $J_2/J_1 = 0.50$.}
    \label{Fig:N3}
\end{figure}

\section{Mean-field results for $N \leq 4$}
\label{Sec:Results2-4}

The implementation of the SCM algorithm needs \textit{a priori} 
knowledge of the periodic structure of possible ground states. 
The results after optimization are sometimes sensitive to the chosen lattice 
geometry, especially for the cluster states that have large unit cells
and break the lattice symmetries. 
To accommodate different candidate ground states as much as possible, 
the two-site unit cell of the primitive honeycomb lattice is enlarged 
along two directions of its lattice vectors by factors $\ell_1$ and $\ell_2$, 
respectively. 
We then consider all enlarged unit cells with the parallelogram geometries 
${\ell_{1,2} \le N}$ with periodic boundary conditions. In our calculation,
the SCM algorithm is not used to handle arbitrary fillings 
currently on a certain unit cell geometry; the number of fermions per cell
is set to be an integer and equal to ${2 \ell_1 \ell_2 / N}$. 
Geometries that do not meet this condition or have any unit width ${\ell_{1,2} = 1}$ 
are excluded. 
Meanwhile the reduced Brillouin zone is discretized into an ${L_1 \times L_2}$ mesh 
with ${L_{1,2} = 50}$ for ${2 \le N \le 6}$ and ${L_{1,2} = 20}$ for ${7 \le N \le 9}$.
For a given $J_2/J_1$ and each allowed geometry, 
the SCM algorithm is run at least 64 times with 
different random seeds to reach the best optimized saddle-point energy 
which is accepted as the global minimum. 
The results are discussed in the following and the ground-state phase diagram 
for ${4 \le N \le 9}$ is presented in Fig.~\ref{Fig:PhaseDiagram}.

We first discuss the numerical results for the SU(2) and SU(3) spins 
as a comparison. For ${N<4}$, the parameter $N$ is not quite a large parameter yet. 
Therefore, the large-$N$ approximation may give incorrect ground states,
and this is what happened in our study of the triangular lattice
where the actual ground state of the SU(2) Heisenberg model is the 
$120^{\circ}$ order~\cite{PhysRevResearch.3.023138}. 
This statement, however, strongly depends on the underlying system.  
For the $J_1$-$J_2$ Heisenberg model on the honeycomb lattice, 
we find the ground state for the SU(2) spins is highly degenerate and 
is given by any dimer covering state on the nearest-neighbor bonds in the range ${0.0 \le J_2/J_1 \le 0.5}$,
and the expectation values of the next-nearest-neighbor bonds are exactly zero.
When ${J_2/J_1 \ge 0.5}$, the ground state is given by  
any dimer covering state on the next-nearest-neighbor bonds, and
the expectation values of nearest-neighbor bonds are exactly zero. 
This is certainly a significant deviation from the actual ground states. 
Nevertheless, the SCM algorithm reproduces the hexagonal plaquette 
VCS state correctly for the SU(3) spins at ${J_2 / J_1 = 0}$ as shown 
in Fig.~\ref{Fig:N3}(a). It is found that the finite interaction $J_2$ 
induces nonvanishing bond 
expectations $\chi_{\braket{\braket{\bm{r}\bm{r}}}}$ that form the three-site simplex 
VCS on two triangular sublattices illustrated in Figs.~\ref{Fig:N3}(b) and~\ref{Fig:N3}(c).
The hexagonal VCS order coexists with the the three-site simplex VCS order 
up to ${J_2 / J_1 = 0.5}$ and then the system smoothly enters into the 
decoupled limit where ${\chi_{\braket{\bm{r}\bm{r}'}} = 0}$.
It is worth noting that the three-site simplex VCS state is still not the 
true ground state for the SU(3) spins on the decoupled triangular 
sublattice~\cite{PhysRevB.85.125116}.
It is expected that the mean-field ground states in the large-$N$ approximation 
become more reliable when the systematic error is suppressed more for 
${N>3}$. 

\begin{figure}[t]
    \centering
    \includegraphics[width=8.5cm]{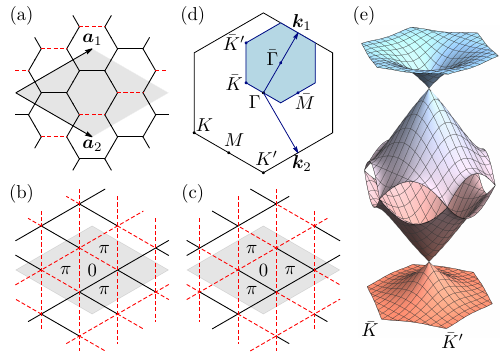}
    \caption{(Color online.) The Dirac spin liquid for the SU(4) spins. 
    The gauge choice implements (a) $\pi$ flux on each hexagonal 
    plaquette formed by the nearest-neighbor bonds and (b, c) 
    staggered $0$/$\pi$ flux     
    on the triangular plaquettes formed by the next-nearest-neighbor bonds. 
    The hopping amplitudes are positive on the solid black bonds and negative on 
    the dashed red bonds. The gray diamonds indicate the same enlarged unit cell 
    containing eight sites. (d) The original (outermost) and reduced (blue hexagon) 
    Brillouin zones and the high-symmetry momenta. (e) The twofold-degenerate spinon 
    spectrum within the reduced Brillouin zone for the Dirac spin liquid here.}
    \label{Fig:DSL}
\end{figure}

\subsection{Dirac spin liquid and its instability to tetramerization}


For the SU(2) antiferromagnetic Heisenberg model on the honeycomb lattice, 
there could be a Dirac node of the spinon bands without any flux at the mean-field level.
Another DSL state has also been proposed for the SU(4) spins where the spinon filling is 
at $1/4$~\cite{PhysRevX.2.041013} and the mean-field theory is characterized by 
\begin{eqnarray}
&&    \chi_{\braket{\bm{r}\bm{r}'}} = |\chi_1|e^{\imath a_{\braket{\bm{r}\bm{r}'}}},\\
&&    \sum_{\bm{r}\bm{r}' \in \text{hex}} a_{\bm{r}\bm{r}'} = \pi, \\
&&     \mu_{\bm{r}} = 0.
\end{eqnarray} 
All the nearest-neighbor bonds have a uniform expectation value $|\chi_1|$ 
but are modulated by a U(1) gauge field $a_{\braket{\bm{r}\bm{r}'}}$ such 
that the gauge flux is equal to $\pi$ per hexagon. 
This is confirmed by our calculation. It is further found that such a $\pi$-flux DSL state 
is stable against the presence of the next-nearest-neighbor interactions 
until ${J_2 / J_1 \approx 0.50}$. 
Specifically, the mean-field saddle 
point gives a compatible gauge flux pattern 
on the triangular sublattices as
\begin{eqnarray}
&&    \chi_{\braket{\braket{\bm{r}\bm{r}'}}} = 
|\chi_2|e^{\imath a_{\braket{\braket{\bm{r}\bm{r}'}}}}, \\ 
  &&  \sum_{\bm{r}\bm{r}' \in \text{tri}} a_{\bm{r}\bm{r}'} = 0 \text{ or }\pi,
\end{eqnarray}
where $|\chi_2|$ is dependent on $J_2/J_1$ and different from $|\chi_1|$. 
The gauge choice $a_{\braket{\braket{\bm{r}\bm{r}'}}}$ on the two types of 
sublattices is not independent once the one on the nearest-neighbor bonds is fixed. 
In Figs.~\ref{Fig:DSL}(a)-~\ref{Fig:DSL}(c), we illustrate the mean-field ansatz refined 
from our numerical results. It turns out that there is a staggered flux 
of zero and $\pi$ on each sublattice, which is nothing but the ansatz of the DSL 
state on the triangular lattice~\cite{PhysRevB.93.144411}. 
Consequently, the Dirac nodes are intact as the ${J_2 / J_1 = 0}$ 
case except for the energy shifts. 
In the reduced Brillouin zone [see Fig.~\ref{Fig:DSL}(d)], 
we plot the spinon band structure around the center $\bar{\Gamma}$. 
The Dirac touchings occur at $\bar{\Gamma}$ and $\bar{K}$ ($\bar{K}'$) 
points for the quarter and half fillings [see Fig.~\ref{Fig:DSL}(e)].

\begin{figure}[t]
    \centering
    \includegraphics[width=8.6cm]{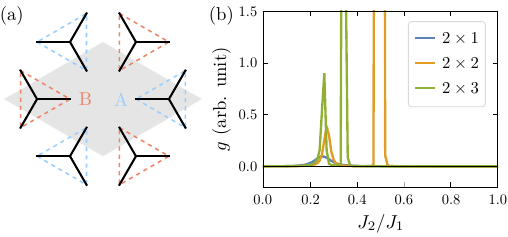}
    \caption{
(Color online.)    
    (a) The ordered singlet tetramers (solid) for SU(4) spins on the nearest-neighbor bonds and two types of three-site simplex VCSs (dashed and colored) around the center of the tetramer on the next-nearest-neighbor bonds. The gray diamond indicates the enlarged unit cell. (b) The fidelity metric $g$ as a function of $J_2/J_1$ for different supercells.}
    \label{Fig:Tetramer}
\end{figure}

With the increasing of the next-nearest-neighbor interactions, a four-site VCS state 
prevails over the DSL state before the decoupled limit at 
${J_2 / J_1 = 1.5}$. The spinons tetramerize as an SU(4) singlet 
on four adjoining sites as sketched in Fig.~\ref{Fig:Tetramer}(a). 
Such a tetramerization was also obtained by the variational Monte Carlo 
approach~\cite{PhysRevB.87.224428}. There are two nonequivalent 
but symmetry-related tetramers within the ${2 \times 2}$ enlarged unit cell. 
At the same time, a three-site simplex VCS forms on the next-nearest-neighbor bonds 
around each center of the tetramers, leaving the center site dangling.  
In the decoupled limit, the dangling site is also incorporated to form 
the SU(4) singlet on the triangular sublattice. 
Here, the four-site VCS state is accompanied by a large degeneracy 
in the large-$N$ limit, because any covering of the tetramers has 
the same energy. Only one type of covering of the tetramers is 
demonstrated here.

\subsection{Fidelity susceptibility}
 
To confirm the stability of the $\pi$-flux DSL state, we further apply the exact diagonalization (ED) method~\cite{KAWAMURA2017180} to different geometries of the enlarged unit cell and calculate the ground-state fidelity metric~\cite{PhysRevB.85.205124} as a function of the next-nearest-neighbor interaction $J_2$, 
\begin{equation}
    g = \frac{2}{N_s} \frac{1 - |\braket{\psi(J_2)|\psi(J_2 + \delta J_2)}|}{(\delta J_2)^2}.
\end{equation}
As only one parameter is varied, this entry of fidelity metric is 
known as fidelity susceptibility.
Here the ground-state wave function $\psi$ is obtained 
by the ED method and $N_s$ is the number of sites within the supercell. 
The fidelity $|\braket{\psi(J_2)|\psi(J_2 + \delta J_2)}|$ measures 
the orthogonality between two ground states that are infinitesimally 
close to each other in the $J_2$ space. The fidelity tends to vanish 
when there is a symmetry-breaking or topological phase transition 
and thus results in a visible peak for the quantity $g$. The latter has 
been treated as a sensitive indicator of Mott transition for the 
SU($N$) Hubbard model~\cite{PhysRevA.84.043601}. 

As shown in Fig.~\ref{Fig:Tetramer}(b), we implement the fidelity 
calculations on supercells with the geometries of ${2 \times 1}$, 
${2 \times 2}$, and ${2 \times 3}$. The number of sites 
${N_s = 4}$, $8$, and $12$ respectively. For all geometries, 
there is a fairly consistent peak at ${J_2/J_1 \approx 0.26}$ 
that becomes sharper with the increasing of system size. 
It could be an evidence for the persistence of the DSL state, although 
the critical value $J_2/J_1$ is about half of the mean-field result.  
There seems to be another phase transition for larger $J_2/J_1$ 
but no consistent results for geometries considered here. 
In fact, the potential four-site VCS state is incompatible with 
the lattice geometries that contain an odd width. 
The failure implies that the finite-size effect is still 
prominent for the stronger next-nearest-neighbor interactions.

\section{Mean-field results for ${5 \le N \le 9}$}
\label{Sec:Results5-9}

Going beyond ${N=4}$, the expanding symmetry group stimulates 
the quantum fluctuations effectively and favors nonmagnetic states 
with a huge classical degeneracy. 
This is believed to be present on any lattice for large enough $N$.
On the honeycomb lattice, the magnetic frustration is incorporated 
by the next-nearest-neighbor Heisenberg interactions for the case
of the SU(2) spins. 
The classical degeneracy may be significantly augmented, resulting in novel quantum states. 
Nevertheless, peculiar quantum-ordered states could also be stabilized due to the competition between the geometric frustration and the structure of the large SU($N$) symmetry.
The synergism and antagonism of the two ingredients 
are further explored for the large-$N$ regime in this section.

\begin{figure}
    \centering
    \includegraphics[width=8.6cm]{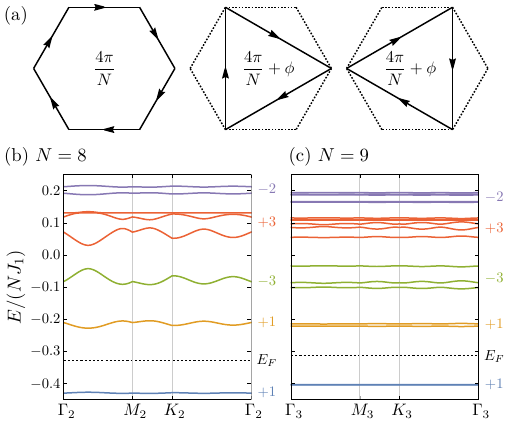}
    \caption{(Color online.) (a) The CSL states for ${5 \le N \le 9}$ resemble the Haldane model 
    with a total flux $4\pi/N$ through the unit cell. The gauge fluxes per hexagonal and triangular plaquettes are indicated by arrows where $\phi$ is only dependent on $J_2$. (b, c) The spinon spectra for CSL states with ${N=8}$ and $9$ at ${J_2/J_1=0.5}$. 
    The spinon bands are grouped in colors according to their separations and labeled with their total first Chern numbers. 
    High-symmetry momenta with the subscript $\ell$ are defined in the reduced hexagonal Brillouin zones for ${\ell \times \ell}$ enlarged unit cells.}
    \label{Fig:CSL}
\end{figure}

\subsection{Chiral spin liquid for ${5 \le N \le 9}$}
\label{Subsec:CSL}

The putative CSL states for SU(${N > 4}$) spins are confirmed in 
our mean-field calculations and are quite stable against the 
next-nearest-neighbor interactions especially for larger $N$ as shown in Fig.~\ref{Fig:PhaseDiagram}. 
The fall of the CSL states occurs at ${J_2/J_1 \approx 1.06}$, 
$0.20$, $0.71$, $1.27$, and $1.59$ for ${5 \le N \le 9}$, respectively. 
This CSL state preserves all lattice translational symmetries 
but breaks the time-reversal and parity symmetries spontaneously 
through the following mean-field ansatz on the nearest-neighbor bonds,
\begin{eqnarray}
&&    \chi_{\braket{\bm{r}\bm{r}'}} = |\chi_1|e^{\imath a_{\braket{\bm{r}\bm{r}'}}},\\
&&    \sum_{\bm{r}\bm{r}' \in \text{hex}} a_{\bm{r}\bm{r}'} = \frac{4\pi}{N}, \\
&&     \mu_{\bm{r}} = 0,
\end{eqnarray}
and on the next-nearest-neighbor bonds, 
\begin{eqnarray}
&&    \chi_{\braket{\braket{\bm{r}\bm{r}'}}} 
= |\chi_2|e^{\imath a_{\braket{\braket{\bm{r}\bm{r}'}}}}, \\ 
&&    \sum_{\bm{r}\bm{r}' \in \text{A;tri}} a_{\bm{r}\bm{r}'} 
= \sum_{\bm{r}\bm{r}' \in \text{B;tri}} a_{\bm{r}\bm{r}'}= \frac{4\pi}{N} + \phi,
\end{eqnarray}
where the U(1) gauge fields $a_{\bm{r}\bm{r}'}$ and 
the corresponding gauge fluxes per hexagonal and   
triangular plaquettes are indicated by arrows in Fig.~\ref{Fig:CSL}(a). 
The letters $A$ and $B$ label two types of triangular sublattices, 
and the U(1) flux difference $\phi$ is dependent on $J_2$ for a given $N$.
It is obvious that such a mean-field solution resembles
the well-known Haldane model~\cite{PhysRevLett.61.2015} 
for the quantum anomalous Hall effect despite that 
the total flux through the unit cell 
is quantized to $4\pi/N$ instead of zero. 
For even parameters $N$ (half-integer spins), 
the CSL state has a spinon spectrum with $N$ bands 
where only the lowest one is fully occupied. 
On the other hand, there are $2N$ bands for odd parameter $N$ (integer spins) 
and the lowest two are fully filled by the spinons.
In Figs.~\ref{Fig:CSL}(b) and~\ref{Fig:CSL}(c), we plot such spinon band structures 
or the CSL states with ${N=8}$ and 9 in their reduced Brillouin zones. 
The corresponding unit cells are enlarged by ${2 \times 2}$ and ${3 \times 3}$, 
respectively. 
Therefore, the reduced Brillouin zones are still hexagons and the high-symmetry momenta are defined similarly as in Fig.~\ref{Fig:DSL}(d). 
For simplicity, we distinguish them by the subscript $\ell$ for an $\ell \times \ell$ enlarged unit cell hereafter.  
For both even and odd parameters $N$, there is a finite energy gap 
between occupied spinon bands and others with higher energies 
due to the U(1) gauge pattern, which means the absence of the spinon Fermi surface. 
Since the spinon bands are not well separated from each other in general, 
the associated first Chern number is generalized to 
\begin{equation}
    C_1=\frac{1}{2\pi}\int_{\text{BZ}} d\bm{k} \Tr \left[\mathcal{F}_{ij} (\bm{k})\right],
\end{equation}
where the non-Abelian Berry curvature 
$\mathcal{F}_{ij} = \partial_i \mathcal{A}_j-\partial_j \mathcal{A}_i - \imath [\mathcal{A}_i, \mathcal{A}_j]$ is written 
in terms of the matrix-valued Berry connection 
$\mathcal{A}_i$ that has the elements 
$[\mathcal{A}_{i}]^{nm} = \imath \braket{\psi_n|\partial_i|\psi_m}$~\cite{Nakahara}. 
Here $(k_i,k_j)$ is the momentum in the Brillouin zone and 
${\partial_i = \partial/\partial_{k_i}}$. The trace is taken over 
the eigenstates denoted as $\psi_n(\bm{k})$ for the $n$th 
energy level. 
In Figs.~\ref{Fig:CSL}(b) and~\ref{Fig:CSL}(c), the spinon bands are also grouped according to the separations and labeled with their total first Chern numbers. 
It turns out that the occupied spinon bands possess the first Chern number ${C_1 = \pm 1}$ 
(the sign is determined by the chirality of the U(1) gauge flux).  
This is a universal result of the CSL phase with different parameters $N$ and interactions. 
Therefore, this gapped system exhibits a nonzero Hall conductivity ${\sigma_{xy}=N/2\pi}$ 
after counting contributions from all $N$ spin flavors. 
By integrating the gapped spinon out, one can further obtain a Chern-Simons
term in the action 
\begin{equation}
\label{Eq:CSterm}
    \mathcal{S}_{\text{CS}} = \frac{N}{4\pi} \int dt d\bm{r}  \epsilon_{\mu\nu\lambda} a_{\mu}\partial_{\nu}a_{\lambda},
\end{equation}
describing the dynamics of U(1) gauge field in $2+1$ dimensions and in the continuum limit. 
It is the Chern-Simons term that determines the topological properties of the CSL state. 
In fact, it endows the spinon with an attached flux $2\pi-2\pi/N$ and renders the fractional statistics to the excitations known as anyons. 
For CSL states found here, the low-energy physics of the U(1) gauge fluctuation can 
be captured by a topological quantum field theory containing 
anyons with a statistical angle $\pi \pm \pi/N$. 
With open boundaries, there are also gapless chiral modes with 
the spin degrees of freedom propagating along the edges of the system, which are described by the chiral SU($N$)$_1$ Wess-Zumino-Witten (WZW) model.


\begin{figure}[t]
    \centering
    \includegraphics[width=8.6cm]{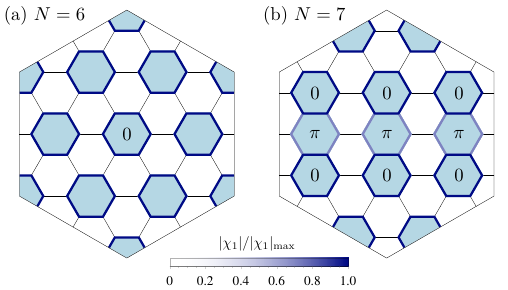}
    \caption{(Color online.) (a) The six-site VCS state for the SU(6) spins. (b) The 14-site VCS state for the SU(7) spins. The bond expectation values $|\chi_1|$ are colored with respect to their relative magnitudes in each case. The shaded plaquettes enclose zero or $\pi$ flux as indicated. $J_2/J_1$ is taken to be $1.00$.}
    \label{Fig:VCS}
\end{figure}

\subsection{Valence cluster solid and stripe states}
\label{Subsec:VCS-Stripe}

In general, the formation of an SU($N$) singlet requires at least $N$ spins. 
These multisite singlets may be rapidly transformed into each other due to the SU($N$) exchanges. 
Nevertheless, as shown in Fig.~\ref{Fig:PhaseDiagram}, two stable VCS states are identified for $N = 6$ and $7$ in the regimes ${0.20 \lesssim J_2/J_1 \lesssim 1.53}$ and ${0.71 \lesssim J_2/J_1 \lesssim 1.63}$, respectively. 
In Fig.~\ref{Fig:VCS}, we present two such VCS states and color the nearest-neighbor bond expectation values $|\chi_1|$ with respect to their relative magnitudes. 
For SU(6) spins, the singlet contains six spins and manifests as a hexagon plaquette with a uniform $|\chi_1|$ and zero flux. 
This is reminiscent of the same structure proposed for SU(3) spins depicted in Fig.~\ref{Fig:N3}(a). 
The SU(7) case is more complicated. 
A 14-site cluster is formed on the nearest-neighbor bonds through two different expectation values and ordered with long periods as shown in Fig.~\ref{Fig:VCS}(b). 
Note that the center hexagon with four weak bond magnitudes encloses $\pi$ flux.

\begin{figure}[t]
    \centering
    \includegraphics[width=8.6cm]{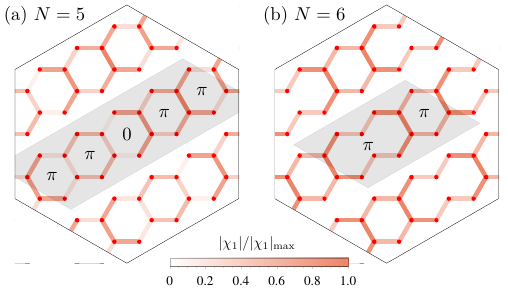}
    \caption{
(Color online.)    
    Stripe state with (a) a $5 \times 2 $ enlarged unit cell for SU(5) spins and (b) a $3 \times 2 $ enlarged unit cell for SU(6) spins. Dashed quadrilaterals indicate the enlarged unit cell. The bond expectation values $|\chi_1|$ are colored with respect to their relative magnitudes in each case. Some regions enclose zero or $\pi$ flux as indicated. $J_2/J_1$ is taken to be $1.60$.}
    \label{Fig:Stripe}
\end{figure}

For the SU(5) and SU(6) spins, the intermediate quantum states manifest as stripe orders in the regimes $1.07 \lesssim J_2/J_1 \lesssim 1.70$ and $1.53 \lesssim J_2/J_1 \lesssim 1.63$, respectively. 
In Fig.~\ref{Fig:Stripe} we illustrate these stripe states on the nearest-neighbor bonds at $J_2/J_1 = 1.60$. 
It can be found that, the inhomogeneous expectation values on the nearest-neighbor bonds lead to a enlarged unit cell containing 20 (12) sites for the $N=5$ ($N=6$) case. 
Both cases have a doubling to one of the primitive honeycomb lattice vectors. 
Along this direction, the bond expectation values vanish alternatively and result in a noncontact stripe pattern. 
This doubling is consistent with the adjacent stripe state of SU(5) spins in the decoupled limit~\cite{PhysRevResearch.3.023138}. 
Along the other direction, the enhanced periodicities are quintuple (triple) for the $N=5$ ($N=6$) case.
Some regions enclose zero or $\pi$ flux as indicated in Fig.~\ref{Fig:Stripe}, but the total flux within the enlarged unit cell is still zero and thus the time-reversal symmetry is preserved. 
The stripe states are also extensively degenerate; several different distributions of bond expectation values are found by the SCM algorithm and they have the exact same stripe pattern including the flux ordering. 
Likewise, the degeneracy is expected to be lifted upon incorporating perturbative $1/N$ corrections.

\subsection{Flux ordered state for ${N=8}$} 
\label{Subsec:FO}

\begin{figure}
    \centering
    \includegraphics[width=8.6cm]{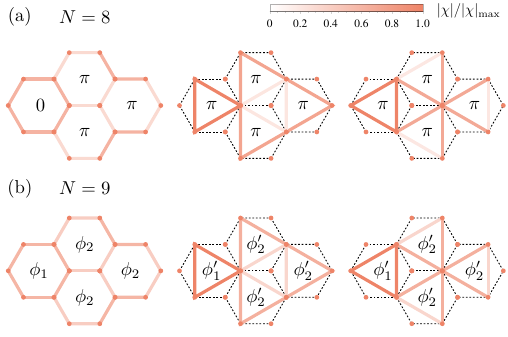}
    \caption{(Color online.) (a) The FO state with a ${2 \times 2}$ enlarged unit cell 
    for ${N=8}$ at ${J_2/J_1=1.50}$. The gauge fluxes are constantly fixed to zero 
    or $\pi$. (b) The ICSL state for ${N=9}$ at ${J_2/J_1=1.70}$. 
    This state also has a ${2 \times 2}$ enlarged unit cell. 
    The total flux on four hexagonal plaquettes is 
    ${\phi_1+3\phi_2=2\pi/9}$. 
    The two fluxes on triangular plaquettes within the same hexagon 
    satisfy the similar relationship as the CSL state. 
    The bond expectations are colored 
    with respect to their relative magnitudes in each case.}
    \label{Fig:ICSLFO}
\end{figure}

Apart from the VCS and stripe states, an intermediate state breaking the lattice translational symmetries is found for SU(8) spins in the range ${1.27 \lesssim J_2/J_1 \lesssim 1.83}$. 
There is a period doubling along both directions of the primitive honeycomb lattice vectors. 
Within the ${2 \times 2}$ enlarged unit cell, the U(1) gauge fluxes per hexagonal and triangular plaquettes are always zero or $\pi$ and ordered in a pattern shown in Fig.~\ref{Fig:ICSLFO}(a). 
Thus,  the lattice translation symmetry is explicitly broken. 
Hence we refer to it as the flux ordered state. 
The time-reversal symmetry is preserved here, and the CSL state does not apply here. 
The spinon band structure consists of eight twofold-degenerate bands 
where the lowest one is fully filled and all others are empty. 
To preserve the point symmetries of the underlying honeycomb lattice, 
we take a supercell with the ${4 \times 4}$ geometry and present 
the spinon band structure in the reduced Brillouin zone as shown 
in Fig.~\ref{Fig:ICSLFOSpectra}(a). Here, each band possesses 
a fourfold degeneracy due to the extra unit cell doubling. 
Although the eight groups of spinon bands are well separated 
from each other, their first Chern number vanishes due to the 
time-reversal symmetry. As the spinon in this flux ordered state 
is fully gapped, the spinon deconfinement is unstable to the U(1) 
gauge fluctuation, and the system will become confined. 
The system would behave more like a confined valence cluster state.

\subsection{Inhomogeneous chiral spin liquid for ${N=9}$}
\label{Subsec:ICSL}

\begin{figure}
    \centering
    \includegraphics[width=8.6cm]{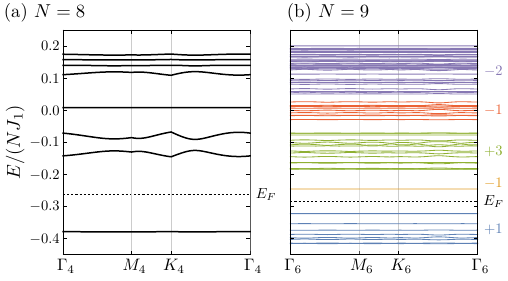}
    \caption{(Color online.) (a) Spinon spectrum for the FO state at ${J_2/J_1=1.50}$. 
    There is a fourfold degeneracy for each band. (b) Spinon spectrum for the ICSL state 
    at ${J_2/J_1=1.70}$. The spinon bands are grouped in colors according to their separations and labeled with their total first Chern numbers. High-symmetry momenta with the subscript $\ell$ are defined in the reduced hexagonal Brillouin zones for ${\ell \times \ell}$ enlarged unit cells.}
    \label{Fig:ICSLFOSpectra}
\end{figure}

Interestingly, there exists another type of CSL state for the SU(9) 
spins in the narrow range ${1.59 \lesssim J_2/J_1 \lesssim 1.75}$.
This state further breaks the lattice translation symmetries     
and hence differs from the CSL states discussed previously. 
The inhomogeneous bond expectation values result in 
a ${2 \times 2}$ enlarged unit cell as shown in Fig.~\ref{Fig:ICSLFO}(b). 
There are two types of gauge flux 
depending on $J_2/J_1$ through four hexagonal plaquettes 
and the total flux within the enlarged unit cell is ${\phi_2+3\phi_2=2\pi/9}$, 
half of that in the homogeneous CSL state. 
A similar ${2 \times 2}$ inhomogeneous CSL state superimposed 
on an average $2\pi/N$ flux per plaquette has also been reported 
as the lowest competing state for the antiferromagnetic SU(5) spins 
on the square lattice~\cite{PhysRevLett.103.135301}. 
Moreover, within each hexagonal plaquette, the gauge fluxes 
piercing two types of the triangular plaquettes satisfy the similar relationship 
of the homogeneous counterpart described in Sec.~\ref{Subsec:CSL}. 
Thus the ICSL state is also a variant of the Haldane model. 
In Fig.~\ref{Fig:ICSLFOSpectra}(b), the spinon band structure is plotted 
in the reduced Brillouin zone corresponding to a ${6 \times 6}$ enlarged unit cell. 
It is clear that the flux pattern results in a spinon band structure with 72 bands, 
where only the lowest eight are fully occupied, and they are separated from the others by a gap.
The associated first Chern numbers for grouped bands are calculated and marked 
in Fig.~\ref{Fig:ICSLFOSpectra}(b). 
In spite of the inhomogeneity, the occupied spinon bands 
also possess the first Chern number ${C_1=\pm 1}$, the same as the CSL states.
Therefore, the ICSL state exhibits a total Hall conductivity ${\sigma_{xy} = N/2\pi}$ 
as well and its effective action contains the same Chern-Simons term as
Eq.~\eqref{Eq:CSterm} in the continuum limit after integrating out the gapped spinons. 
The obtained low-energy effective theory is a topological quantum field theory 
with the chiral Abelian topological order and anyonic statistics. 
The spinon is converted into anyons with a statistical angle ${\pi \pm \pi/N}$ 
in the analogous manner of the homogeneous case. 
It is expected that gapless chiral states carrying spin degrees of freedom are 
supported by the ICSL as edge modes, and their effective theory 
is also described by the SU($N$)$_1$ WZW model.

\section{Discussion}
\label{Sec:Conclusion}

In this work, we perform a mean-field analysis on the SU($N$) Heisenberg 
model on the honeycomb lattice with both nearest- and next-nearest-neighbor 
antiferromagnetic interactions. In the large-$N$ approximation, 
a variety of intermediate ground states are identified 
subject to the strict local constraints. For the SU(4) spins, the DSL state with a gauge flux 
$\pi$ per hexagonal plaquette and its instability towards the tetramerized spin singlets 
when the next-nearest-neighbor exchange interaction predominates are confirmed. 
The putative CSL states at ${J_2 = 0}$ for higher SU($N$) are also obtained and 
the associated gauge fluxes per hexagonal plaquette follow the same form of $4\pi/N$. 
The next-nearest-neighbor bond expectations would develop in the presence of 
$J_2$ and render the ground state to a variant of the Haldane model with $4\pi/N$ 
total flux through the unit cell. These CSL states preserve the lattice translation 
 and have a unity first Chern number for the occupied spinon bands. 
A series of intermediate quantum phases breaking various lattice translation 
symmetries would appear when the next-nearest-neighbor interactions become 
stronger. For the SU(8) and SU(9) spins, inhomogeneous states with ${2 \times 2}$ 
ordering pattern are found. While the former preserves the time-reversal symmetry, 
the latter is identified as a new ICSL state. Apart from the VCS states containing 
6 sites (or 14 sites) found for the SU(6) [or the SU(7)] spins, there are two types 
of stripe ordered states manifesting themselves in the form of a doubled period 
along one of the honeycomb lattice vectors.

The SU(4) DSL state is supposed to be the ground state of the spin-orbital SU(4) 
symmetric Kugel-Khomskii model of Mott insulators on the honeycomb lattice. 
There might be an intrinsic instability due to the monopole proliferation. 
There is no definitive conclusion for the stability of the SU(4) DSL state at this stage.   
Our mean-field results provide an evidence for its stability against the 
antiferromagnetic next-nearest-neighbor interaction. 
This is also supported by the fidelity analysis using the ED method and consistent 
with previous variational Monte Carlo study~\cite{PhysRevB.87.224428}. 
But all numerical methods suffer the significant finite-size effect in the nearby tetramerized state, resulting in an undetermined phase boundary. The mean-field phase boundary obtained in this work is  overestimated and would be modified by the $1/N$ corrections.
More numerical tools such as tensor network algorithms are needed 
to give a conclusive result.
On the other hand, multispin interaction terms e.g., 
the scalar spin chirality, could destabilize the DSL state in principle. 
The consequence on current results is an open question.

\begin{acknowledgments}
    We thank Chun-Jiong Huang for useful discussions.
    This work is supported by the National Science 
Foundation of China with Grant No. 92065203, by the Ministry of Science and Technology 
of China with Grants No. 2018YFE0103200 and No. 2021YFA1400300, by the Shanghai Municipal Science and Technology Major Project with Grant No. 2019SHZDZX04, and by the Research Grants 
Council of Hong Kong with General Research Fund Grant No. 17306520. 
\end{acknowledgments}

\appendix

\section{The SCM algorithm}
\label{appendix}

The SCM algorithm is a nondeterministic optimization algorithm starting from the randomly initialized auxiliary fields $\chi_{\bm{r}\bm{r}'} = |\chi_{\bm{r}\bm{r}'}| e^{\imath \varphi_{\bm{r}\bm{r}'}}$  and unified chemical potentials $\mu_{\bm{r}}$ on a given geometry~\cite{PhysRevB.84.174441}. 
For convenience, both the amplitudes and phases of the bond operators $\chi_{\bm{r}\bm{r}'}$ are taken to be uniform distributions, e.g., $|\chi_{\bm{r}\bm{r}'}| \in [0.02, 0.20]$ and $\varphi_{\bm{r}\bm{r}'} \in [0, 2\pi]$, respectively. 
The default value of $\mu_{\bm{r}}$ is chosen to be zero before optimization. 
It is obvious that the random initial state violates the desired single occupation in general and one can denote the deviation of the local fermion density as
\begin{equation}\label{eq:density_deviation}
    \delta n_{\bm{r}} = 1 - \braket{f_{\bm{r}\alpha}^{\dagger}f_{\bm{r}\alpha}},
\end{equation}
where the expectation value of the local density operator is obtained using the ground state of the mean-field Hamiltonian $\mathcal{H}_{\text{MF}}$ with initialized parameters at this stage. 
To correct the density, one must adjust the chemical potential $\mu_{\bm{r}}$ locally by $\delta \mu_{\bm{r}}$. 
As we will show later, this adjustment will be implemented iteratively to achieve the self-consistency; therefore, it is sufficient to consider the lowest order at each step, which can be expressed as 
\begin{equation}\label{eq:Green_function}
    \delta \mu = \mathcal{G}^{-1} \delta n.
\end{equation}
Here $\delta \mu$ and $\delta n$ are column vectors with elements $\delta \mu_{\bm{r}}$ and $\delta n_{\bm{r}}$, respectively. 
The response matrix $\mathcal{G}$ is nothing but the inverse of the density-density correlation in real space and at zero frequency. 
It is a real symmetric matrix by definition. 
In principle, the correlation elements  $\mathcal{G}_{\bm{r}\bm{r}'}$ can be calculated from the noninteracting mean-field Hamiltonian $\mathcal{H}_{\text{MF}}$. 
So far, all derivations are done within the framework of the standard linear response theory. 
However,there is a caveat that the invertibility of the density-density correlation cannot be guaranteed despite the fact that all its eigenvalues are real. 
Actually, at least one eigenvalue of the matrix $\mathcal{G}$ must be exactly zero because it is trivial to adjust the chemical potentials uniformly at each site. 
The fermion density will remain intact and result in a naive divergence of $\mathcal{G}^{-1}$. 
To remediate this fatal flaw, Hermele and Gurarie proposed a modified diagonalization procedure in Ref.~\cite{PhysRevB.84.174441}. 
In particular, only nonzero eigenvalues $g_i$ are focused on after diagonalizing 
\begin{equation}
    \mathcal{G} = U g U^{-1}.
\end{equation}
We have assigned the index $i$ to the diagonalized basis. 
In such a basis, the linear transformation in Eq.~\eqref{eq:Green_function} formally reduces to 
\begin{equation}
    U^{-1} \delta \mu = g^{-1} U^{-1}\delta n.
\end{equation}
By mapping the vanishing eigenvalues to infinity, the adjustment of the chemical potential $\delta \mu $ becomes well defined; that is, 
\begin{equation}
    (U^{-1} \delta \mu)_i  = 
    \begin{cases}
        g^{-1}_i U^{-1}\delta n, & g_i \neq 0 \\
        0, & g_i=0.
    \end{cases}
\end{equation}

With the above relationship, a new mean-field Hamiltonian $\mathcal{H}_{\text{MF}}$ and related ground state can be generated by a simple replacement, $\mu_{\bm{r}} \rightarrow \mu_{\bm{r}} + \delta \mu_{\bm{r}}$. 
The local fermion density $\delta n_{\bm{r}}$ should be updated concurrently, resulting in a new deviation $\delta n_{\bm{r}}$ as Eq.~\eqref{eq:density_deviation}.  
Then, the problem returns to find the adjustment of chemical potentials in the current ground state. 
These processes complete the self-consistent procedure. 
The problem of searching for an appropriate set of chemical potential deviation $\delta \mu_{\bm{r}}$ can be solved by iterating the procedure until a fixed point is reached. 
This is the core of the SCM algorithm to strictly impose the local constraints $n_{\bm{r}}=\braket{f_{\bm{r}\alpha}^{\dagger}f_{\bm{r}\alpha}}=1$. 

After meeting the first saddle-point condition, Eq.~\eqref{eq:local_constraint}, the others about the bound operators still need to be satisfied. 
With the modified chemical potentials calculated in the previous stage, an updated set of auxiliary fields $\chi_{\bm{r}\bm{r}'}$ can be determined via 
\begin{align}
    & \chi_{\braket{ \bm{r}\bm{r}' }} \rightarrow 
    - \braket{ f_{\bm{r}'\alpha}^{\dagger} f_{\bm{r}\alpha}^{} }/N, \\ 
    &  \chi_{\braket{\braket{ \bm{r},\bm{r}' }}} \rightarrow  - J_2/J_1 \braket{ f_{\bm{r}'\alpha}^{\dagger} f_{\bm{r}\alpha}^{} }/N. 
\end{align}
Once again, the local constraints are violated if the system has not yet converged to the true saddle point. 
The amended auxiliary fields, together with the chemical potentials satisfying the single occupation condition in the previous stage, can be treated as a new and better starting point. 
The two-stage updating procedure is thus implemented iteratively until reaching a convergence in energy within a given numerical error. 
It has been proved rigorously in Ref.~\cite{PhysRevB.84.174441} that within each optimization process the energy of the final state must be less than or equal to that of the initial state. 
Therefore, the SCM algorithm ends at one of the local energy minima. 
In order to best estimate the global minimum, the program should be run with differently initialized fields, which can be realized by varying the random number seeds. 
Finally, a collection of local minima will be reaped and the lowest one is accepted as the best result of the ground state. 

\bibliography{SUNRef.bib}

\end{document}